\documentclass[letterpaper,12pt]{article}\usepackage[]{graphicx}\usepackage[]{color}
\makeatletter
\def\maxwidth{ %
  \ifdim\Gin@nat@width>\linewidth
    \linewidth
  \else
    \Gin@nat@width
  \fi
}
\makeatother

\definecolor{fgcolor}{rgb}{0.345, 0.345, 0.345}

\usepackage{framed}
\makeatletter
 {\par\unskip\endMakeFramed%
 \at@end@of@kframe}
\makeatother

\definecolor{shadecolor}{rgb}{.97, .97, .97}
\definecolor{messagecolor}{rgb}{0, 0, 0}
\definecolor{warningcolor}{rgb}{1, 0, 1}
\definecolor{errorcolor}{rgb}{1, 0, 0}
\usepackage{alltt}

\usepackage{setspace}
\usepackage{graphicx}
\usepackage{times}
\usepackage{hyperref}
\usepackage{xcolor}
\definecolor{blue}{rgb}{0.1,0.1,0.65}
\definecolor{green}{rgb}{0.0,0.5,0.25}
\hypersetup{
    colorlinks, urlcolor={blue}, citecolor={green},
}

\hypersetup{pdfpagemode=UseNone} % don't show bookmarks on initial view

\usepackage{titlesec}
\titleformat*{\section}{\sffamily \bfseries}
\IfFileExists{upquote.sty}{\usepackage{upquote}}{}

\begin{document}
\setstretch{2.0}

\begin{center}
{\sffamily Experience report}

\vspace{50mm}

\large
\textbf{Fourteen years of R/qtl: Just barely
  sustainable}

\vspace{10mm}

Karl W. Broman

Department of Biostatistics \& Medical Informatics

University of Wisconsin--Madison
\end{center}

\vfill

\setstretch{1.2}
\hfill
\begin{minipage}{3in}
\textbf{Address for correspondence}:
\bigskip

\qquad \begin{tabular}{ll}
\multicolumn{2}{l}{Karl W Broman} \\
\multicolumn{2}{l}{Department of Biostatistics \& Medical Informatics} \\
\multicolumn{2}{l}{University of Wisconsin-Madison} \\
\multicolumn{2}{l}{2126 Genetics/Biotechnology Center} \\
\multicolumn{2}{l}{425 Henry Mall} \\
\multicolumn{2}{l}{Madison, Wisconsin 53706} \\
\textbf{Phone}: & 608-262-4633 \\
\textbf{Email}: & kbroman@biostat.wisc.edu
\end{tabular}

\end{minipage}
\setstretch{2.0}

\clearpage

\section*{Abstract}

R/qtl is an R package for mapping quantitative trait loci (genetic
loci that contribute to variation in quantitative traits) in
experimental crosses. Its development began in 2000.  There have been
38 software releases since 2001.  The latest release contains
35k lines of R code and
24k lines of C code, plus
15k lines of code for the documentation.
Challenges in the development and maintenance of the
software are discussed.
A key to the success of R/qtl is
that it remains a central tool for the chief developer's own research
work, and so its maintenance is of selfish importance.

\clearpage

\section*{Introduction}

If two inbred strains (for example, of plants) show consistent differences in a
quantitative trait, one can be confident
that the difference is genetic. An experimental cross between the two
strains can be used to identify the genetic loci (called quantitative
trait loci, QTL) that contribute to the trait difference: we seek
genomic regions for which genotypes are associated with the trait.

As an illustration, Figure~1 displays the results of QTL analysis with
data from \cite{Moore2013}, on gravitropism in Arabidopsis.
LOD curves measuring the strength of association between
phenotype and genotype are displayed in Figure~1A; a plot of
phenotype vs.\ genotype at the marker exhibiting the strongest
association is shown in Figure~1B.

% Figure 1
\begin{figure*}[tbh]
\begin{center}

\includegraphics[width=\textwidth]{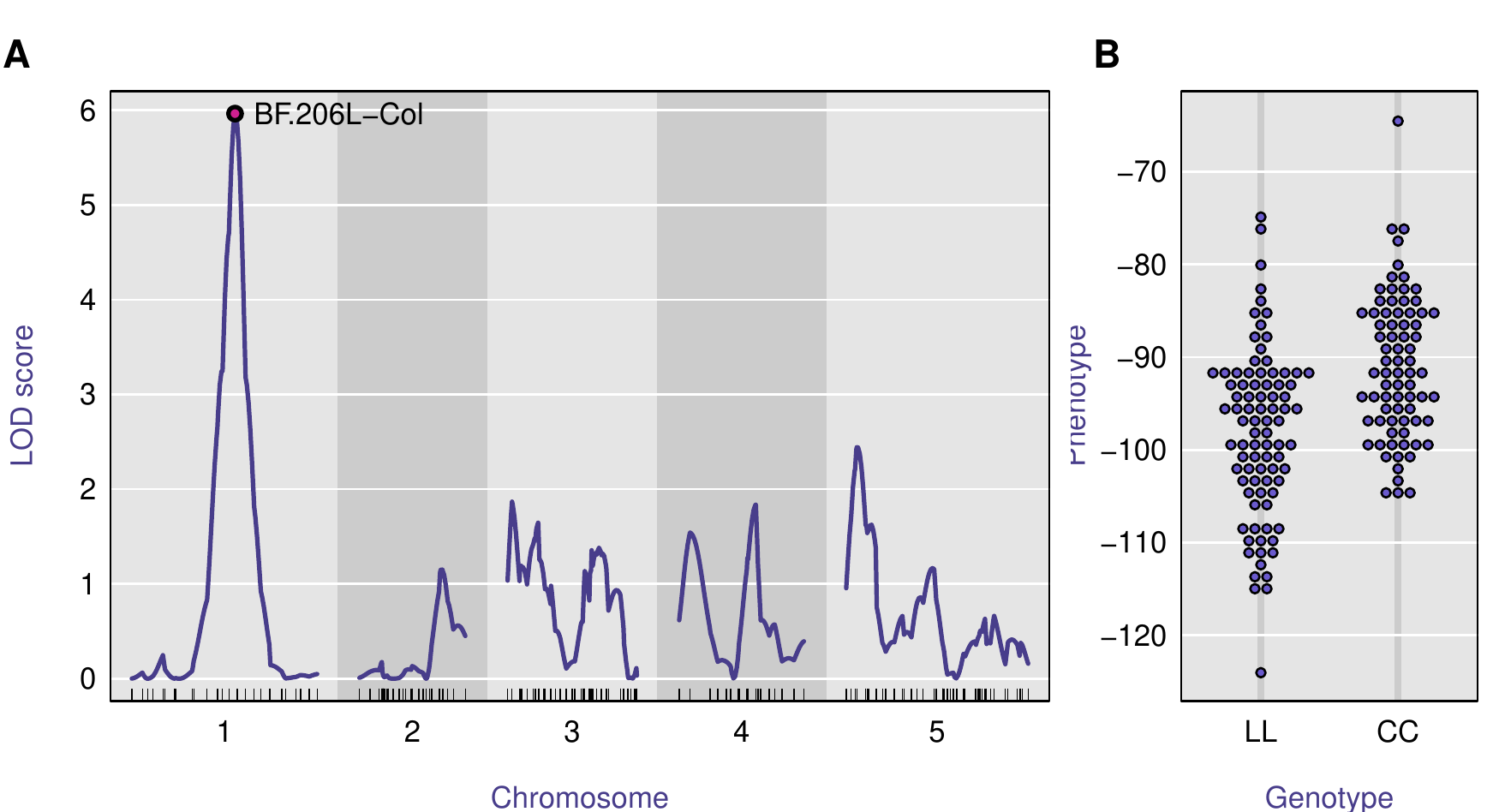}

\caption{Typical analysis results from R/qtl. \textbf{A}: LOD curves
  across the genome, measuring association between phenotype and
  genotype across, and \textbf{B}: Association between genotype and
  phenotype at the marker with the strongest association. The data
  are from \cite{Moore2013}; panel B was created using the R package
  \href{http://www.cbs.dtu.dk/~eklund/beeswarm/}{beeswarm}.
 }
\end{center}
\end{figure*}
\newpage

Numerous software packages for QTL analysis are available, some
commercial (e.g.,
\href{http://www.kyazma.nl/index.php/mc.MapQTL}{MapQTL} and
\href{http://www.multiqtl.com}{MultiQTL}) and some free
(e.g. \href{http://www.broadinstitute.org/ftp/distribution/software/mapmaker3}{Mapmaker/QTL},
and \href{http://statgen.ncsu.edu/qtlcart/index.php}{QTL
Cartographer}).

My own QTL mapping software, \href{http://www.rqtl.org}{R/qtl} \cite{rqtlpaper},
is developed as an add-on package to the widely used
general statistical software, \href{http://www.r-project.org}{R}
\cite{RCoreTeam2013}. The software is open source, licensed under
GPL3, and currently hosted on GitHub.

\section*{History}

I became interested in QTL mapping in graduate school, twenty years ago.
Mark Neff introduced me to Lander and Botstein's
paper on interval mapping \cite{Lander1989}, which remains
the most commonly used QTL analysis method.

In the fall of 1999, I joined the Department of Biostatistics at Johns
Hopkins University as an assistant professor. In February, 2000, Gary
Churchill visited me from the Jackson Laboratory, and we discussed our
shared interest in QTL analysis methods and the need for more advanced
software. Gary suggested that we write our own QTL mapping package. He
was thinking Matlab, but I was keen to use R. (R version 1.0 was
released the following week.) R won out over Matlab largely because I
developed a working prototype more quickly. I had recently written
some C code implementing the hidden Markov model (HMM) technology
\cite{Baum1970} for the treatment of missing genotype information for QTL
analysis in experimental crosses.  This served as the starting point
for the package.

Our main goal was for the software to enable the exploration of
multiple-QTL models. We also wanted it to be easily extendible as new
methods were developed.  My initial concept was to implement all QTL
mapping methods, good and bad, so that their performance could be
compared within a single package.

Much of the development of R/qtl occurred during a three-month
sabbatical I spent at the Jackson Laboratory in Fall, 2001.  (It is
easy to remember the year, because I was there on 11 September 2001.)  Hao
Wu, a software engineer working with Gary from 2001--2005, contributed
a great deal to the core of R/qtl. In 2009--2010, Danny Arends (with
some assistance from Pjotr Prins and me) incorporated Ritsert Jansen's
MQM code \cite{Jansen1993, Jansen1994, JansenStam1994}, previously
available only in commercial software, as part of R/qtl \cite{Arends2010}.

\section*{About me}

I am an applied statistician. My primary interest is in helping other
scientists answer questions using data. But that generally requires
the development of new statistical methods, and such methods must be
implemented in software.  Thus, I spend a considerable amount of time
programming.

I have little formal training in programming or software engineering,
and I am not a specialist in computational statistics.  But I think
it's important for an applied statistician to be self-sufficient: We
can't rely on others to develop the tools we need but must be able to
do that ourselves.

\section*{Strengths}

R/qtl has a number of strengths. It is comprehensive: It includes
implementations of many QTL mapping methods, and it has a number of
tools for the fit and exploration of multiple-QTL models. It has
extensive facilities for data diagnostics and visualizations. It can
be extended, as the results of important intermediate calculations,
that serve as the basis for any QTL mapping method, are exposed to the
user.

The central calculations are coded in C, for speed, but R is used for
manipulating data objects and for graphics. Figure~2 displays the
growth of code in R/qtl over time, as the number lines of code in R
and C, as well as in the R documentation files. Ignoring the
documentation files, the code is about
60\% R and
40\% C.

\begin{figure*}[tbh]
\begin{center}

\includegraphics[width=\textwidth]{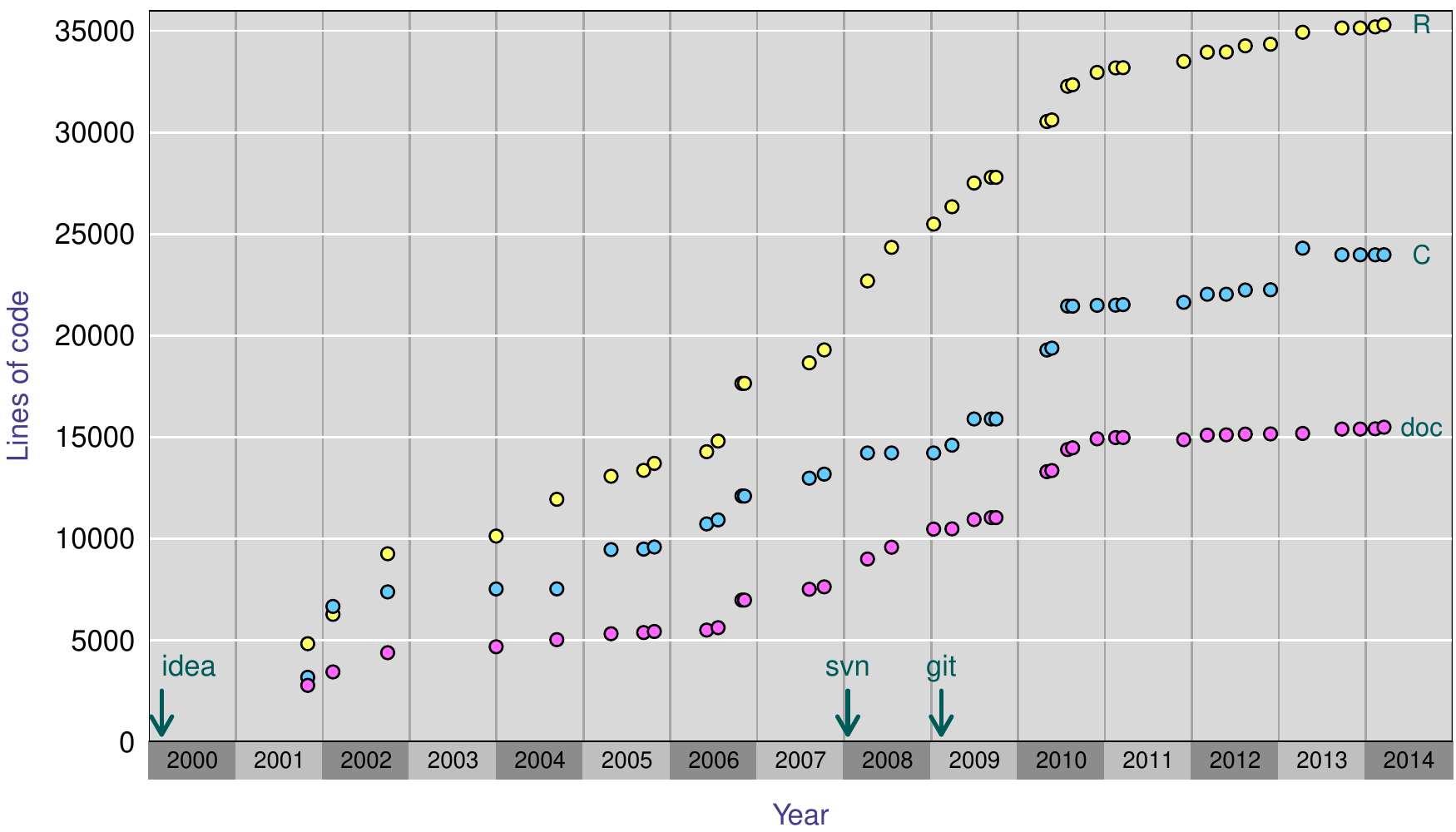}

\caption{Numbers of lines of code in released versions of R/qtl over
  time. (Yellow and blue correspond to R and C code, respectively;
  pink is for the R documentation files.)}
\end{center}
\end{figure*}

Developing the software as a package for R has a number of advantages
for the developer, particularly to make use of R's facilities for
graphics and input/output, as well as its extensive math/stat software
library. The R packaging system and the
Comprehensive R Archive Network (CRAN, \url{http://cran.r-project.org})
simplifies the software installation process and makes it easy to
provide documentation.

The user also benefits by having the QTL mapping software embedded
within general statistical software: for further data analysis before
and after the QTL analysis, and for developing specially-tailored
visualizations. R also provides an excellent interactive environment
for data analysis.

A number of related packages are coordinated with R/qtl, using a
common data structure or input file format and some shared functions.
Examples include \href{http://www.ssg.uab.edu/qtlbim/}{qtlbim}
\cite{Yandell2007},
\href{http://cran.r-project.org/web/packages/wgaim/index.html}{wgaim}
\cite{Taylor2011},
and \href{http://cran.r-project.org/web/packages/dlmap/}{dlmap}
\cite{Huang2012}.

\section*{Weaknesses}

R/qtl also has a number of weaknesses. There has largely been one
developer (who doubles as the support staff), and so many desired features have not
been implemented. We never wrote formal specifications for the
internal data formats, which makes it more difficult for others to
contribute code to the project.

There have been a number of memory management issues, particularly
regarding the copying of large datasets in memory as it is moved from
R to C. This can reduce performance in extremely high-dimensional
applications.

The biggest weakness is that the central data structure is too
restrictive. This makes it difficult to extend the software beyond
the simplest types of experimental crosses.

\section*{Some really bad code}

More embarrassing than the above weaknesses is that, while R/qtl
contains some good code (like the HMM implementation), there is also
some really terrible code. I've learned a great deal about programming
while developing R/qtl, but it's hard to go back and replace old code.
(And some of this code was obviously bad when written and just should
have been constructed with greater care.)

The worst piece of code (now fixed: see the commit in the git
repository on GitHub,
\url{https://github.com/kbroman/qtl/commit/4cd486})
involved a
really stupid approach to find the first non-blank element in a
vector of character strings. The code worked fine in small datasets,
and so it took a while to discover the problem. (Open source means
everyone can see my stupid mistakes. Version control means everyone
can see every stupid mistake I've ever made.)

In many cases, functions aren't split up into short reusable functions
the way they should be, and there is a lot of repeated code.  And some
of the functions have very long lists of arguments, which can be
really intimidating to users.

In lots of cases, bugs were fixed by adding little band-aids of code.
Functions get longer and longer to handle more and more special cases,
to the point that they are nearly unreadable.  These very long,
complex functions are a barrier to the further development of the
software.

The worst offender is {\tt scantwo()},
for performing a two-dimensional genome scan for pairs of QTL.  The R
function is 1402 lines long (that is
4\% of the R code in the
package). And the R code is just moving data and results around.  The
actual calculations are performed in C, in a series of files that
comprise 4725 lines, which is
20\% of the C code in the package.

\section*{Version control}

In the first eight years of developing R/qtl, I didn't use formal
version control.  Initially, I was simply editing the code in place
and saving copies of releases when I sent them to the R Archive.
Incorporating
others' contributions was often a hassle.

In the fall of 2006, \'Saunak Sen and I used a
Subversion repository to
facilitate the development of our book about QTL mapping and R/qtl
\cite{rqtlbook}, but it wasn't until January, 2008, that I began
to use subversion for R/qtl, as well. In February, 2009, Pjotr Prins
helped me to switch to git and to place the
R/qtl repository on GitHub.

I'm not sure how I managed the project before adopting version
control. Version control makes it so easy to try things out without
fear of breaking working code. And collaboration on software development is
terribly tedious and inefficient without version control.

\section*{User support}

Helping users to solve their research problems can be quite rewarding.
But it can also require considerable patience. I respond to several queries per
week about R/qtl, mainly through a Google Group
that was
started in September, 2004, but many others directly via email.

I had hoped that the
R/qtl discussion
group would become a network of folks answering each others'
questions about R/qtl and QTL mapping, but I am basically the only one
answering questions. I think that's partly because, to avoid spam, I
moderate all messages, at which point I generally try to answer
them. Thus, participants see my answer at basically the same time as
they see the original question. But at least these electronic
conversations are public and searchable.

I've helped many people revise their data files for use with R/qtl.
It's very hard to predict what might go wrong in data import and to
provide appropriate error messages.

Questions are often frustrating. They may provide too little detail
(``I tried such-and-such but \emph{it didn't work}.'') or too much
(``Could you please look at the attached 25-page Word document
containing code and output and tell me if I'm doing something wrong?'')
Many questions are not so much about the software but are more general
scientific or data analysis questions.

Many questions would be answered by a careful reading of
the documentation, but software documentation is often dreadfully
boring. I've learned that the most popular documentation are the short
tutorials with practical examples clearly illustrating important
tools. These are time-consuming to create (and maintain).

Some things are \emph{possible\/} in R/qtl but are not well documented
and not really feasible except for users with considerable programming
and analysis experience. It feels wrong to say, ``That \emph{is\/}
possible, but I don't have the time to explain how.''

It was helpful to have written a book about QTL mapping and R/qtl
\cite{rqtlbook}, but it's frustrating to watch the publisher
nearly double its street price. If I could do it again, I would
self-publish.

\section*{Sustainable academic software}

Why put so much effort into software development? The principal
advantage is the software itself: QTL analysis is the focus of much of
my own research, and software that is easy for others to use is also
easy for me to use. Second, R/qtl provides a platform for me to
distribute implementations of QTL mapping methods that I develop, such
as the two-part model of \cite{Broman2003} and the model
selection methods of \cite{Broman2002} and \cite{Manichaikul2009}.
Third, the software has led to many interesting
consulting-type interactions and a good number of collaborations (leading
to co-authorship on papers and some grant support). Fourth, my own
methods grant is much more attractive with a successful software aim.
Finally, the software supports others' research, and my primary
duty as a statistician and an academic is to help others.

I think the key to the sustainability of piece of scientific software
is that the developer continues to use the software
for his/her own research. So often, the developer moves on to other
research problems, and the software is orphaned.

But software development requires considerable time and
support. Often, researchers don't recognize the many indirect benefits
of building useful software tools, nor the long-term benefits of
devoting extra time to the construction of \emph{good\/} software. The
reviewers of grant proposals may recognize the value of software
development, but generally it can not be the sole research effort, and
funding agencies often over-emphasize innovation, which can make it
difficult to obtain support for software maintenance. Moreover,
academic culture encourages researchers to create independent software
packages rather than contribute to existing ones.

The biggest open question, for me personally, is how to encourage the
formation of a community around a given software development effort:
for coding, for answering users' questions, and for developing
improved documentation and tutorials.

\section*{Future}

As Fred Brooks said in \emph{The Mythical Man Month},
``Plan to throw one away; you will anyhow.'' \cite{Brooks1975},
Ch.\ 11) This was
restated by Eric Raymond in \emph{The Cathedral and The Bazaar}:
``You
often don't really understand the problem until after the first time
you implement a solution. The second time, maybe you know enough to do
it right.'' \cite{Raymond1999}

In collaboration with Danny Arends and Pjotr Prins, I've initiated a
reimplementation of R/qtl, with a focus on high-dimensional data and
more modern cross designs, such as the Collaborative Cross \cite{CTC2004}
and the Mouse Diversity Outbred Population \cite{Svenson2012}.
We will also
focus on interactive graphical tools, implemented with the
Javascript library, \href{http://d3js.org/}{D3}.

There has been a renewal of interest in QTL analysis, particularly
with the growth of eQTL analysis, in which genome-wide gene expression
measures are treated as phenotypes
(see, for example, \cite{Jansen2001,Tesson2009}).
We hope that our new software will become
a popular platform for the analysis of large-scale eQTL data, as R/qtl
has been for more traditional-sized QTL projects.

\section*{Acknowledgments}

This work was supported in part by NIH grant R01 GM074244.
Christina Kendziorski generously provided comments to improve the manuscript.
Numerous
people have contributed to R/qtl over the years: Danny Arends, Gary
Churchill, Ritsert Jansen, Steffen Moeller, Pjotr Prins, \'Saunak Sen,
Laura Shannon, Hao Wu, and Brian Yandell.

\clearpage

\bibliographystyle{vancouver}
\bibliography{rqtlexper}

\begin{thebibliography}{10}

\bibitem{Moore2013}
Moore CR, Johnson LS, Kwak IY, Livny M, Broman KW, Spalding EP.
\newblock High-Throughput Computer Vision Introduces the Time Axis to a
  Quantitative Trait Map of a Plant Growth Response.
\newblock Genetics. 2013;195:1077--1086.
\newblock Available from:
  \url{http://www.ncbi.nlm.nih.gov/pmc/articles/PMC3813838/}.

\bibitem{rqtlpaper}
Broman KW, Wu H, Sen S, Churchill GA.
\newblock {R}/qtl: {QTL} mapping in experimental crosses.
\newblock Bioinformatics. 2003;19:889--890.
\newblock Available from:
  \url{http://bioinformatics.oxfordjournals.org/content/19/7/889.long}.

\bibitem{RCoreTeam2013}
{R Core Team}.
\newblock R: A language and environment for statistical computing; 2013.
\newblock Available from: \url{http://www.r-project.org}.

\bibitem{Lander1989}
Lander ES, Botstein D.
\newblock Mapping {M}endelian factors underlying quantitative traits using
  {RFLP} linkage maps.
\newblock Genetics. 1989;121:185--199.
\newblock Available from:
  \url{http://www.ncbi.nlm.nih.gov/pmc/articles/PMC1203601/}.

\bibitem{Baum1970}
Baum LE, Petrie T, Soules G, Weiss N.
\newblock A maximization technique occurring in the statistical analysis of
  probabilistic functions of Markov chains.
\newblock Ann Math Statist. 1970;41:164--171.
\newblock Available from: \url{http://www.jstor.org/stable/2239727}.

\bibitem{Jansen1993}
Jansen RC.
\newblock Interval mapping of multiple quantitative trait loci.
\newblock Genetics. 1993;135:205--211.
\newblock Available from:
  \url{http://www.ncbi.nlm.nih.gov/pmc/articles/PMC1205619/}.

\bibitem{Jansen1994}
Jansen RC.
\newblock Controlling the type {I} and type {II} errors in mapping quantitative
  trait loci.
\newblock Genetics. 1994;138:871--881.
\newblock Available from:
  \url{http://www.ncbi.nlm.nih.gov/pmc/articles/PMC1206235/}.

\bibitem{JansenStam1994}
Jansen RC, Stam P.
\newblock High resolution of quantitative traits into multiple loci via
  interval mapping.
\newblock Genetics. 1994;136:1447--1455.
\newblock Available from:
  \url{http://www.ncbi.nlm.nih.gov/pmc/articles/PMC1205923/}.

\bibitem{Arends2010}
Arends D, Prins P, Jansen RC, Broman KW.
\newblock {R}/qtl: high throughput multiple {QTL} mapping.
\newblock Bioinformatics. 2010;26:2990--2992.
\newblock Available from:
  \url{http://www.ncbi.nlm.nih.gov/pmc/articles/PMC2982156/}.

\bibitem{Yandell2007}
Yandell BS, Mehta T, Banerjee S, Shriner D, Venkataraman R, Moon JY, et~al.
\newblock {R}/qtlbim: {QTL} with {B}ayesian interval mapping in experimental
  crosses.
\newblock Bioinformatics. 2007;23:641--643.
\newblock Available from:
  \url{http://bioinformatics.oxfordjournals.org/content/23/5/641.long}.

\bibitem{Taylor2011}
Taylor J, Verbyla A.
\newblock R package wgaim: {QTL} analysis in bi-parental populations using
  linear mixed models.
\newblock J Statist Soft. 2011;40:1--18.
\newblock Available from: \url{http://www.jstatsoft.org/v40/i07/paper}.

\bibitem{Huang2012}
Huang BE, Shah R, George AW.
\newblock dlmap: An {R} package for mixed model {QTL} and association analysis.
\newblock J Statist Soft. 2012;50:1--22.
\newblock Available from: \url{http://www.jstatsoft.org/v50/i06/paper}.

\bibitem{rqtlbook}
Broman KW, Sen S.
\newblock A guide to {QTL} mapping with {R}/qtl.
\newblock Springer; 2009.

\bibitem{Broman2003}
Broman KW.
\newblock Mapping quantitative trait loci in the case of a spike in the
  phenotype distribution.
\newblock Genetics. 2003;163:1169--1175.
\newblock Available from:
  \url{http://www.ncbi.nlm.nih.gov/pmc/articles/PMC1462498/}.

\bibitem{Broman2002}
Broman KW, Speed TP.
\newblock A model selection approach for the identification of quantitative
  trait loci in experimental crosses.
\newblock J R Statist Soc B. 2002;64:641--656.
\newblock Available from: \url{http://www.jstor.org/stable/3088807}.

\bibitem{Manichaikul2009}
Manichaikul A, Moon JY, Sen S, Yandell BS, Broman KW.
\newblock A model selection approach for the identification of quantitative
  trait loci in experimental crosses, allowing epistasis.
\newblock Genetics. 2009;181:1077--86.
\newblock Available from:
  \url{http://www.ncbi.nlm.nih.gov/pmc/articles/PMC2651044/}.

\bibitem{Brooks1975}
Brooks FP.
\newblock Mythical Man Month: Essays on Software Engineering.
\newblock Addison-Wesley; 1975.

\bibitem{Raymond1999}
Raymond ES.
\newblock The Cathedral \& the Bazaar.
\newblock O'Reilly, Sebastopol, CA; 1999.
\newblock Available from:
  \url{http://www.catb.org/~esr/writings/cathedral-bazaar/cathedral-bazaar/}.

\bibitem{CTC2004}
{The Complex Trait Consortium}.
\newblock The {C}ollaborative {C}ross, a community resource for the genetic
  analysis of complex traits.
\newblock Nat Genet. 2004;36:1133--1137.
\newblock Available from:
  \url{http://www.nature.com/ng/journal/v36/n11/full/ng1104-1133.html}.

\bibitem{Svenson2012}
Svenson KL, Gatti DM, Valdar W, Welsh CE, Cheng R, Chesler EJ, et~al.
\newblock High-resolution genetic mapping using the {M}ouse {D}iversity outbred
  population.
\newblock Genetics. 2012;190:437--447.
\newblock Available from:
  \url{http://www.ncbi.nlm.nih.gov/pmc/articles/PMC3276626/}.

\bibitem{Jansen2001}
Jansen RC, Nap JP.
\newblock Genetical genomics: the added value from segregation.
\newblock Trends Genet. 2001;17:388--391.
\newblock Available from:
  \url{http://www.cell.com/trends/genetics/abstract/S0168-9525(01)02310-1}.

\bibitem{Tesson2009}
Tesson BM, Jansen RC.
\newblock {eQTL} analysis in mice and rats.
\newblock In: DiPetrillo K, editor. Cardiovascular Genomics, Methods in
  Molecular Biology 573. 10. Humana Press; 2009. p. 285--309.

\end{thebibliography}

\end{document}